\documentclass[sn-mathphys,Numbered]{sn-jnl}% Math and Physical Sciences Reference Style
\usepackage{graphicx}%
\usepackage{multirow}%
\usepackage{amsmath,amssymb,amsfonts}%
\usepackage{amsthm,bm}%
\usepackage{mathrsfs}%
\usepackage[title]{appendix}%
\usepackage{xcolor}%
\usepackage{textcomp}%
\usepackage{manyfoot}%
\usepackage{booktabs}%
\usepackage{algorithm2e}
\usepackage{algorithmic}%
\RestyleAlgo{ruled}
\usepackage{listings}%
\usepackage{cleveref}
\usepackage{caption}
\usepackage{subcaption}
\theoremstyle{thmstyleone}%
\theoremstyle{thmstyletwo}%
\theoremstyle{thmstylethree}%

\begin{document}

\title[Article Title]{Improvement of Heatbath Algorithm in LFT using Generative models}


\author[1]{\fnm{Ali} \sur{Faraz}}\email{ali21579@gmail.com}

\author[3,4]{\fnm{Ankur} \sur{Singha}}\email{a.singha@tu-berlin.de}

\author[2]{\fnm{Dipankar} \sur{Chakrabarti}}\email{dipankar@iitk.ac.in}
\author[3,4,5]{\fnm{Shinichi}\sur{Nakajima}}
\author[1]{\fnm{Vipul}\sur{Arora}}\email{vipular@iitk.ac.in}

\affil[1]{\orgdiv{Department of Electrical Engineering}, \orgname{IIT Kanpur}, \orgaddress{\postcode{208016}, \state{Uttar Pradesh}, \country{India}}}

\affil[2]{\orgdiv{Department of Physics}, \orgname{IIT Kanpur}, \orgaddress{\postcode{208016}, \state{Uttar Pradesh}, \country{India}}}
\affil[3]{Probabilistic Modeling and Inference Group, BIFOLD,
Germany}
\affil[4]{Machine Learning Group, Technische Universit{\"a}t Berlin,
Germany}

\affil[5]{RIKEN Center for AIP,Japan}
% \affil[3]{\orgdiv{Department}, \orgname{Organization}, \orgaddress{\street{Street}, \city{City}, \postcode{610101}, \state{State}, \country{Country}}}


\abstract{
The Heatbath Algorithm is commonly used for sampling in local lattice field theories, but performing exact updates or sampling from the local density is challenging when dealing with continuous variables. Heatbath methods rely on rejection-based sampling at each site, which can suffer from low acceptance rates if the proposal distribution is not optimally chosen—a non-trivial task. In this work, we propose a novel, straightforward approach for generating proposals at each lattice site for the $\phi^4$ and XY models using generative AI models. This method learns a conditional local distribution, without requiring training samples from the target, conditioned on both neighboring sites and action parameter values.
%Conditional generative models are emerging as promising tools for sampling in critical regions. When conditioned on the parameters of the distribution, they can be efficaciously extended to critical regions, without the need for retraining. 
%Since the local distributions are simpler, the model is not affected by mode collapse, which generally occurs while training with RKLD. We optimize the model with reverse Kullback-Leibler divergence (RKLD) to avoid the need for ground truth samples. 
}

\keywords{Markov Chain Monte Carlo, Metropolis-Hastings, Lattice models, Normalizing flows, Generative machine learning}

\maketitle
\section{Introduction}
Lattice models in physics describe systems on a discrete grid, where configurations follow a Boltzmann distribution characterized by a Hamiltonian \( H(\phi,\lambda) \) or action \( s(\phi,\lambda) \). The statistical properties of these lattices vary with parameters \( \lambda \), exhibiting significant changes near specific values of \( \lambda \) known as critical regions, where phenomena like critical slowing down occur. In these regions, traditional sampling methods like MCMC, while providing convergence guarantees, become inefficient due to high autocorrelation times. Specialized algorithms, such as Swendsen-Wang \cite{SwendsenWang}, Wolff \cite{wolff1989comparison,wolff1990} address this but have limitations, particularly for continuous symmetries. The Hamiltonian Monte Carlo (HMC) \cite{neal_hmc}, which makes global MCMC updates, is considered the state of the art for lattice QCD simulations.  Recently, generative machine learning (ML) methods such as normalizing flows \cite{rezende2016variationalinferencenormalizingflows} and diffusion models \cite{ho2020denoisingdiffusionprobabilisticmodels} have shown promise in efficiently sampling lattice field theories, with successful applications in $\phi^4$ \cite{albergoflowbased,Albergo:2022qfi,Hackett:2021idh,Kim_Nicoli_Paper2,Caselle:2023mvh,Singha:2023cql,Singha:2021nht,pawlowski2020reducing} and gauge theory \cite{Kanwar_2020,Eichhorn:2023uge,Abbott:2022zhs,Abbott:2022hkm,Abbott:2023thq,Wang:2023exq,Singha:2023xxq,Finkenrath:2024pR,zhu2024diffusionmodelslatticegauge,Caselle:2024uen}. Also, recent ML based, AIS variant sampler such as Non-Equilibrium Transport Sampler \cite{Albergo:2024trn} showed successful  application on Gaussian mixture distributions, and statistical lattice field theory. These generative methods however aim to model the joint distribution of the entire lattice, which poses a significant scalability challenge for learning-based approaches. In lattice community a well known approach for sampling local lattice theories is the Heat Bath algorithm, which factorizes the lattice distribution, allowing each lattice site to be sampled 
%based 
conditioned
on its local neighboring sites. This approach is particularly efficient for discrete systems like the Ising model, where only two possible states exist. However, in continuous systems, this requires problem specific designs \cite{PhysRevD.21.2308,KENNEDY1985393} and often requiring rejection-based methods. The difficulty of finding effective proposal distributions for a generic continuous systems can lead to high rejection rates in Heatbath, increasing the simulation cost. Moreover, at different regime of the action parameter, one may need to fine tune the proposal distribution.

We present the Parallelizable Block Metropolis-within-Gibbs (PBMG) method for sampling in local lattice models, offering an efficient proposal distribution for rejection-based Heatbath algorithms. PBMG learns the local single-site distribution conditioned on neighboring sites and the action parameter values. Once trained, the model can be used to sample the entire lattice as in the Heatbath method. The target distribution is a simple one-dimensional distribution, thus significantly improves the learning efficiency. Notably, PBMG does not require any training samples from the target distribution. With a well-trained model, we can reduce the rejection rate commonly encountered in Heatbath sampling. The PBMG operates within a Metropolis-within-Gibbs framework, using ML model-generated proposals conditioned on neighboring sites as well as action parameters. Thus PBMG are conditional generative models, e.g., conditional GMMs and conditional NFs. 

% , and . By focusing on this simplified target distribution, we can also mitigate the issue of mode collapse commonly encountered when sampling from the joint distribution.

To validate the proposed approach, we apply it to 2-D lattices, namely, the XY model from statistical Physics and the scalar $\phi^4$ model from Lattice Field Theory.

\section{Heatbath as Parallelizable Block Metropolis-within-Gibbs (PBMG)}

In this section, we introduce the Heat Bath methods within a generative model framework, which correspond to a Metropolis-within-Gibbs approach, using clear notation.

Consider an $N$-dimensional probability distribution $p(\phi_1, \phi_2, \dots,  \phi_N)$. For a lattice, $N$ is the number of lattice sites and $\phi_i$ is the random variable at site $i$. 
We partition these sites into $G$ partitions such that the distribution of a site $i$ in a partition $g$, conditioned on all sites $j\notin g$, is independent of all sites $i' \in g \setminus i$. 

For implementing MCMC in general state-spaces, one requires to construct  a Markov chain transition kernel $p(\phi_i|\phi_{j\neq i})$ that keeps the target distribution $p(\phi_1, \phi_2, \dots,  \phi_N)$ invariant, and is ergodic for this distribution. Such kernels can also be combined via composition. Keeping this in mind, let $K_i$ be a transition kernel that updates the site $i\in g$, keeping all other sites of the lattice the same. Then the combined kernel that changes all the sites in the partition $g$ is
\begin{equation}
    K_g = \prod_{i\in g} K_i
\end{equation}
and the overall kernel for updating all the sites in a lattice is
\begin{equation}
    K = \prod_g K_g\,.
\end{equation}
 The advantage of partitioning is that all the sites in the same partition can be sampled simultaneously, thereby making the process faster. Moreover, each kernel $K_i$ need not be conditioned on all the sites outside the partition $g$, but only a small number of sites in a local neighbourhood of the site $i$.

Every site-kernel $K_i$ for each $i \in g$ and for every partition $g$ is a Metropolis-within-Gibbs kernel, which means that each site-kernel $K_i$ is a Gibbs kernel with a Metropolis-Hastings accept-reject step.

Let $\bm{\phi}_{-g}$ denote the set of random variables at all the lattice sites excluding the ones in the partition $g$ and $\bm{\psi}$ denote given lattice parameters (e.g., temperature, coupling parameters).
%Define $\phi'$.
Let us define $q(\phi^{(t)}_i | \bm{\phi}_{-g},\bm{\psi} ; \bm{\theta})$ as the parametric proposal distribution, parameterized by $\bm{\theta}$ and conditioned on $\bm{\phi}_{-g}$ and $\bm{\psi}$.
% Here, $\phi^{(t)}_i$ denotes the value of the random variable $\phi_i$ at timestep $t$.
Then, the acceptance probability $\alpha_{K_i}$ for any site-kernel $K_i$ $\forall$ $i \in g$ is
\begin{align}
    \alpha_{K_i} &= \frac{p(\phi^{(t+1)}_i | \bm{\phi}_{-g}, \bm{\psi})}{p(\phi^{(t)}_{i} | \bm{\phi}_{-g}, \bm{\psi})} \cdot \frac{q(\phi^{(t)}_{i} | \phi^{(t+1)}_{i}, \bm{\phi}_{-g}, \bm{\psi} ; \bm{\theta})}{q(\phi^{(t+1)}_{i} | \phi^{(t)}_{i}, \bm{\phi}_{-g}, \bm{\psi} ; \bm{\theta})}
\end{align}
We design a proposal such that,
\begin{align}
    q(\phi^{(t+1)}_i|\phi^{(t)}_i,\bm{\phi}_{-g},\bm{\psi} ; \bm{\theta})=q(\phi^{(t+1)}_i|\bm{\phi}_{-g},\bm{\psi} ; \bm{\theta})
\end{align}
% i.e., a proposal such that the sampling of the corresponding component is independent of the previous value of the component. Such a proposal is called an independent proposal w.r.t. a component or a conditionally independent proposal. Furthermore, we aim to obtain a high acceptance rate from the proposal. 
% The choice of independence proposal here is informed in the sense that it is trained to closely match the conditional distribution as target, thereby leading to improved performance. 
% A conditionally independent proposal with a high acceptance rate would ensure that consecutive samples are very much different from each other which would indirectly lower the autocorrelation within the chain.

% In such a case, the acceptance rate is a direct measure of the performance of the proposal. But it is important to note here that, since, the proposal is an independent proposal w.r.t. a component and not a completely independent proposal, a high acceptance rate cannot guarantee that the integrated autocorrelation time for lattice functions (i.e. functions that take in a lattice as input and output a real value) w.r.t. a Markov chain will be low. This is because the calculation of the integrated autocorrelation time in these cases involves all the lattice points and not just a single component. 

 The maximum acceptance rate possible i.e., an acceptance rate equal to 1 will be achieved when $q(\phi^{(t)}_{i} | \bm{\phi}_{-g}, \bm{\psi} ; \bm{\theta})$ is exactly the same as $p(\phi^{(t)}_{i} | \bm{\phi}_{-g}, \bm{\psi})$. This reduces our goal to design (or learn) a proposal that could sample from the true conditional distribution as closely as possible. In order to achieve this goal, we use methods like Normalizing Flows and Gaussian Mixture Models in generative machine learning. In the next two sections, we apply the PBMG method to the XY model and the Scalar $\phi^4$ theory in 2D.
% In each section, we first analyze the true conditional distribution for a single component in the corresponding PDF. Subsequently, we discuss the structure of the corresponding proposal distribution, the details of the training procedure and the inference procedure.

\section{Application to the XY Model}
% \subsection{Target distribution}
% The local Hamiltonian for the XY model is
% \begin{align}
%     H(\bm{\phi}) = &- \frac{1}{2} \sum_{i,j}\big[ \cos (\phi_{i,j} - \phi_{i+1,j}) + \cos (\phi_{i,j} - \phi_{i,j+1}) \notag  \\ 
%     &+ \cos (\phi_{i,j} - \phi_{i-1,j}) + \cos (\phi_{i,j} - \phi_{i,j-1})\big]
% \end{align}
% Here, $\phi_{i,j}$ is the angular random variable with range $[0,2\pi)$ at the lattice site with coordinates $(i,j)$. Here, we have used periodic boundary conditions for the lattice. The lattice distribution at a given temperature $T\in\mathbb{R}$ is
% \begin{equation}
%     p(\bm{\phi};T) \propto e^{-\frac{H(\bm{\phi})}{T}}
% \end{equation}
The local Hamiltonian of the the XY model for $(i,j)$th component $\phi_{i,j}$ of the lattice vector $\bm{\phi}$ is
\begin{align}
    H(\phi_{i,j}) = &- \big[ \cos (\phi_{i,j} - \phi_{i+1,j}) + \cos (\phi_{i,j} - \phi_{i,j+1}) \notag  \\ 
    &+ \cos (\phi_{i,j} - \phi_{i-1,j}) + \cos (\phi_{i,j} - \phi_{i,j-1})\big]
\end{align}
We see that the Hamiltonian of the $(i,j)$th component depends only on the components of the four nearest neighbours denoted by $n(i,j) = \{(i+1,j), (i,j+1), (i-1,j), (i,j-1)\}$. Therefore the conditional distribution of $\phi_{i,j}$ given the four nearest neighbour components and temperature is 
\begin{equation}
    p\left( \phi_{i,j}|\{\phi_{l,m}: (l,m)\in n(i,j)\}, T \right)=p( \phi_{i,j}|\mathbf{v}_{i,j}) \propto e^{-\frac{H(\phi_{i,j})}{T}}
\end{equation}
%The random variable $\theta_i\in[0,2\pi)$ has a circular topology.
The above conditional distribution is our target distribution. Here, $\mathbf{v}_{i,j} = (\phi_{i+1,j}, \phi_{i,j+1}, \phi_{i-1,j}, \phi_{i,j-1}, T)$ is the $5$x$1$ condition vector corresponding to the site $(i,j)$ which consists of the four nearest neighbour components and the temperature. For this model, we have divided the lattice into two partitions $g_0$ and $g_1$.
\begin{equation}
    g_k = \{(i,j): (i+j)\%2=k\}; k=0,1
\end{equation}
% \begin{figure*}[h]
% \centering
% \includegraphics[scale=0.9]{figures/inference_figure.png}
% \caption {Partitioning used for a lattice in the XY model. Here, the colors white and black represent the two partitions.}
% \label{partitioning}
% \end{figure*}

% \subsection{Modeling the Proposal distribution}
% \begin{figure*}[h]
% \centering
% \includegraphics[scale=0.4]{figures/new_prop_dist.png}
% \caption {Proposal distribution for the XY model}
% \label{6.1}
% \end{figure*}
We use Normalizing Flows to model the proposal distribution $q(\phi_{i,j} | \mathbf{v}_{i,j} ; \bm{\theta})$. 
Using the change of variables formula,
\begin{align}
    \label{eq.11}
    q(\phi_{i,j} | \mathbf{v}_{i,j} ; \bm{\theta})= p_Z \left( f^{-1}(\phi_{i,j} ; \bm{\theta}_R)|\mathbf{v}_{i,j};\bm{\theta}_B \right) \left| \det{\left( \frac{\partial f^{-1}(\phi_{i,j} ; \bm{\theta}_R)}{\partial \phi_{i,j}} \right)} \right|
\end{align}
where, $p_Z(z | \mathbf{v}_{i,j}; \bm{\theta}_B)$ is the base distribution and $f(z ; \bm{\theta}_R)$ is the invertible transformation used in the Normalizing Flow. Here, $\bm{\theta} = \{\bm{\theta_B}, \bm{\theta_R}\}$. We use Rational Quadratic Splines (RQS) as the transform $f$. 

The loss function used in the training procedure is the expected value of the KL divergence between the proposal $q(\phi_{i,j}|\mathbf{v}_{i,j};\bm{\theta})$ and the target $p(\phi_{i,j}|\mathbf{v}_{i,j})$. The Monte Carlo approximation can be used to estimate the KL divergence as follows
\begin{multline}
    \mathcal{L} \approx \frac{1}{n} \cdot \frac{1}{N} \sum_{r=1}^n \sum_{k=1}^N [\log{p_Z(z_k|(\mathbf{v}_{i,j})_r;\bm{\theta}_B)} + \log{|\det{J_f(z_k| (\mathbf{v}_{i,j})_r;\bm{\theta}_R)}|^{-1}} \\
    - \log{p \left(f(z_k ; \bm{\theta}_R) | (\mathbf{v}_{i,j})_r \right)} ]
\end{multline}
For training this model we generate training data $\mathbf{v}_{i,j}$  from $p_v(\mathbf{v}_{i,j})$ i.e.  $\mathbf{v}_{i,j}$ from $p_v(\mathbf{v}_{i,j}) = $ Unif$([0,2\pi]^4 \times [T_1, T_2])$, where $T_2-T_1$ is the training range for temperatures.  For further details on the proposal model, model architecture and the training/inference process for XY model, please refer to Appendix \ref{phi4}.

\section{Application to the $\phi^4$ Theory}
% The lattice $\phi^4$ theory on a 2D lattice has undergone extensive investigation in both statistical mechanics and quantum field theory due to its intricate phase structure and critical behaviour. It demonstrates second-order phase transitions, characterized by alterations in system properties like magnetization or correlation length with varying coupling parameters. For a thorough comprehension of the lattice $\phi^4$ model, one can refer to the works \cite{AKDe,AKDe2, Vierhaus2010Simulation}. In the following section, we will discuss the proposed ML-based sampling approach for the lattice $\phi^4$ theory.
% \subsection{Target Distribution}
% The Euclidean action for scalar $\phi^4$ theory in 2D can be written as,
% \begin{align}
% S(\bm{\phi},\lambda,m^2)=\sum_{i,j}\Bigl(m^2 + 4 \Bigr)\phi^2_{i,j} - \phi_{i,j}[\phi_{i+1,j} + \phi_{i,j+1} + \phi_{i-1,j} + \phi_{i,j-1}]+  \lambda \phi_{i,j}^4 
% \end{align}
%   where $(i,j)$ represents the indices of a lattice site and $\phi_{i,j}$ is a real-valued random variable defined for every lattice site.

% The lattice distribution is given by the Boltzmann distribution law,
% \begin{equation}
%     p({\bm{\phi}|\lambda, m^2}) \propto e^{-S(\bm{\phi}, \lambda, m^2)}
% \end{equation}

The local action for $\phi^4$ theory for lattice site  $(i,j)$ can be written as

\begin{align}
    S_{loc}(\phi_{i,j},\lambda,m^2, \{\phi_{l,m}: (l,m)\in n(i,j)\}) &= S_{loc}(\phi_{i,j},\lambda,m^2,\kappa_{i,j}) \\
    &= \Bigl(m^2 + 4 \Bigr)\phi_{i,j}^2+ \lambda \phi^4_{i,j}-2\phi_{i,j}\kappa_{i,j} 
\end{align}

where $\kappa_{i,j} = \phi_{i+1,j} +\phi_{i,j+1} + \phi_{i-1,j} + \phi_{i,j-1}$.

% We see that the local action for a lattice site depends on its four nearest neighbours i.e. the fields at each lattice site only interact with its nearby lattice points.
The conditional distribution of the lattice site $(i,j)$ can be written as 
\begin{align}
    p\big(\phi_{i,j}|\lambda, m^2, \kappa_{i,j}) = p\big(\phi_{i,j}|\mathbf{v}_{i,j})\propto e^{- S_{loc}(\phi_{i,j},\mathbf{v}_{i,j})}
    \label{targ_phi4}
\end{align}

where, $\mathbf{v}_{i,j}=(\lambda, m^2, \kappa_{i,j})$ is the condition vector for the distribution. For $\phi^4$ theory as well, we have divided the lattice into the same two partitions $g_0$ and $g_1$, where
\begin{equation}
    g_k = \{(i,j): (i+j)\%2=k\}, k=0,1
\end{equation}

% \subsection{Modeling the Proposal Distribution: PBMG-\texorpdfstring{$\phi^4$}{ } }

We construct the proposal distribution for the scalar $\phi^4$ theory by using a Gaussian Mixture Model with six Gaussian components. 
The proposal distribution parameterized by $\bm{\theta}$, can be written as 
\begin{align}
 q(\phi_{i,j}|\mathbf{v}_{i,j}; \bm{\theta}) = \sum_{k=1}^6 \pi_k(\mathbf{v}_{i,j}; \bm{\theta}_k) \mathcal{N}(\phi_{i,j}| \mu_k(\mathbf{v}_{i,j}; \bm{\theta}_k), \sigma_k(\mathbf{v}_{i,j}; \bm{\theta}_k))
\end{align}
 where $\mu_k$, $\sigma_k$ $\pi_k$ are the mean, standard deviation and mixing coefficients of the $k^{th}$ Gaussian distribution, and $\bm{\theta} = \{\bm{\theta}_k \}_{k = 1}^{6}$.

The training procedure of PBMG-$\phi^4$ is similar to that of PBMG-XY. The loss function used in the training procedure is the expected value of the KL divergence between the proposal $q(\phi_{i,j}|\mathbf{v}_{i,j};\bm{\theta})$ and the target $p(\phi_{i,j}|\mathbf{v}_{i,j})$ 

\begin{equation}
    \mathcal{L} \approx \frac{1}{n} \sum_{r=1}^n \left[ \frac{1}{N}\sum_{k=1}^N\left[\log q((\phi_{i,j})_k|(\mathbf{v}_{i,j})_r; \bm{\theta}) - \log p\big((\phi_{i,j})_k|(\mathbf{v}_{i,j})_r)\right] + \|\bm{\pi}\left((\mathbf{v}_{i,j})_r; \bm{\theta} \right) \|\right]
    \label{loss_phi}
\end{equation}

For further details on the architecture and the training/inference process for $\phi^4$ theory, please refer to Appendix \ref{phi4}.
% We sample from $p_v(\mathbf{v}_{i,j})$ to generate 17,500 samples of $\mathbf{v}_{i,j}$ and use this as the training set. We use the Adam optimizer with a learning rate of 0.0001 and default hyperparameters. We perform validation by calculating the average acceptance rate for 50 random sets of parameters ($\lambda$, $m^2$) where $\lambda \in [2.5, 15]$ and $m^2 \in [-8,0]$, and stop training when we achieve an average acceptance rate of around 98\%. PBMG-$\phi^4$ also gets trained very quickly taking only a few minutes to train on a low-end single GPU machine.

% (make corrections)
% To propose a new field value at the lattice site x, we pass the corresponding conditional vector to the trained GMM. The new value is accepted through the MH accept-reject test. To update the whole lattice,  we first update the black sites together and then update the white sites.
% The procedure for MCMC sampling using PBMG-$\phi^4$ is exactly the same as that of PBMG-XY given in Algorithm-\ref{alg1}. During inference, however, while calculating $V_{g_k}$ where $k = 0,1$, we take the absolute value of $\kappa_{i,j}$ and sample $\phi_{i,j}$ from the resulting GMM. We then consider $-\phi_{i,j}$ as the final sampled value and proceed with inference. This procedure is correct because,
% \begin{equation}
%     S_{loc}(\phi_{i,j},\lambda,m^2,-\kappa_{i,j}) = S_{loc}(-\phi_{i,j},\lambda,m^2,\kappa_{i,j})
% \end{equation}
\section{Results for PBMG}
In this section, we assess the performance of the PBMG model in comparison to the heatbath method for both the $\phi^4$ and XY models in 2D. For the $\phi^4$ model, we use the standard heatbath method, where samples are drawn from a Gaussian distribution, followed by a rejection step to account for the $\phi^4$ interaction terms. In the case of the XY model, we use a uniform distribution as the proposal for the rejection step. We compute and compare the acceptance rates between the PBMG and heatbath algorithms. Note that the acceptance rate with rejection sampling differs from that in Metropolis-within-Gibbs sampling. To establish equivalency, we define the acceptance rate for Heatbath algorithm in the following manner.

If $n_i^{(k)}$ represents the number of trials needed to obtain one successful update at site $i$ for the $k$th configuration in the ensemble, then the total number of attempts $A$ for the entire configuration is given by
\[ A_k =\sum_{i=1}^{N} n_i^{(k)} \]
 where $N$ is the total number of lattice sites. The total number of successful updates for a configurations is equal to N.

We define the acceptance rate $R$ for the $k$th configuration as:  $R_k = \frac{\mbox{success updates}}{\mbox{total attempts}} = \frac{N}{A_k}$

For PBMG, the acceptance rate is straightforward to compute, as it simply corresponds to the Metropolis-Hastings acceptance rate.

\begin{figure}[h!]
\centering
    \begin{subfigure}[b]{.49\linewidth} 
        \centering
        \includegraphics[width=\textwidth]{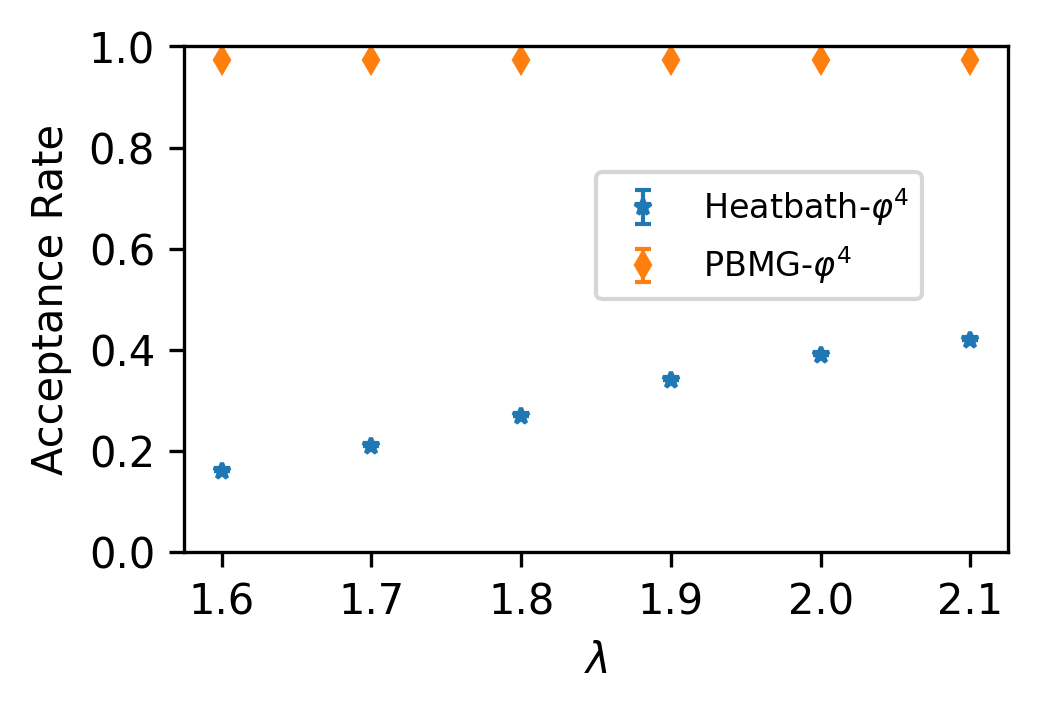}
        \caption{}\label{acc_a}
    \end{subfigure}
    \begin{subfigure}[b]{.49\linewidth} 
        \centering
        \includegraphics[width=\textwidth]{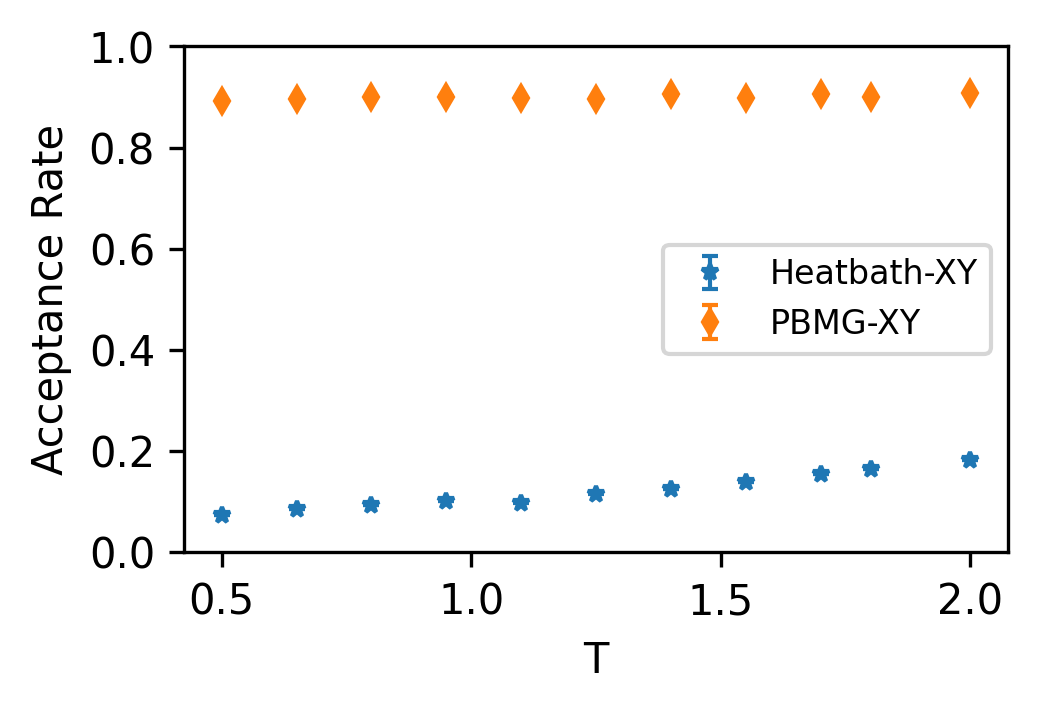}
        \caption{}\label{acc_b}
    \end{subfigure}
\caption{The acceptance rate for both PBMG and Heatbath Algorithm, for lattice size of $64\times64$; a) Phi4 Theory, b) XY Model.} \label{Acceptance}
\end{figure}

The acceptance rate directly reflects the computational cost of our simulation algorithm. In Figure \ref{Acceptance}, we compare the acceptance rates of both methods for the XY model and $\phi^4$ theory.

For the $\phi^4$ theory, we generate samples using both the PBMG and heatbath algorithms for $\lambda$ values in the range $(1.6, 2.1)$, covering both sides of the phase transition. As shown in Figure \ref{acc_a}, the acceptance rate is approximately $98\%$. For the XY model, we compare the acceptance rate across different temperature values in the range $(0.5, 2.0)$, as illustrated in Figure \ref{acc_b}. We observe that the acceptance rate for the PBMG model is close to $90\%$. In both the XY and $\phi^4$ models, the acceptance rate remains nearly constant across the parameter ($T/\lambda$) values.

\section{Conclusion}
We have proposed a generative-based heatbath sampler for local lattice systems with continuous degrees of freedom. Our model, PBMG, is straightforward to train as it learns a one-dimensional distribution and can serve as a proposal for a heatbath sampler. A key advantage of this approach is its conditioning on neighboring sites and the action or Hamiltonian parameters, without requiring any training samples from the target distribution. Unlike traditional methods that may need different proposals for different ranges of the action parameter, the PBMG model offers a flexible, single proposal mechanism for updating all lattice sites while covering a broad range of action parameter values. Extending this improved heatbath approach to gauge theory could be an exciting direction for future research.

\section*{acknowledgment}
This work was supported by Core Research Grant of SERB, Govt. of India (project no. 003466).
AS and SN were supported by the German Federal Ministry of Education and Research under the grant BIFOLD24B.

\appendix

\section*{Appendix A: PBMG for XY model }
\label{XY}
\subsection*{Modeling the Proposal distribution}
% \begin{figure*}[h]
% \centering
% \includegraphics[scale=0.4]{new_prop_dist.png}
% \caption {Proposal distribution for the XY model}
% \label{6.1}
% \end{figure*}
We use Normalizing Flows to model the proposal distribution $q(\phi_{i,j} | \mathbf{v}_{i,j} ; \bm{\theta})$. $p_Z(z | \mathbf{v}_{i,j}; \bm{\theta}_B)$ is the base distribution and $f(z ; \bm{\theta}_R)$ is the invertible transformation used in the Normalizing Flow. Here, $\bm{\theta} = \{\bm{\theta_B}, \bm{\theta_R}\}$. The condition vector that is input to all the neural networks is $\mathbf{v}_{i,j}$. 

Using the change of variables formula,
\begin{align}
    \label{eq.11}
    q(\phi_{i,j} | \mathbf{v}_{i,j} ; \bm{\theta})= p_Z \left( f^{-1}(\phi_{i,j} ; \bm{\theta}_R)|\mathbf{v}_{i,j};\bm{\theta}_B \right) \left| \det{\left( \frac{\partial f^{-1}(\phi_{i,j} ; \bm{\theta}_R)}{\partial \phi_{i,j}} \right)} \right|
\end{align}
We use Rational Quadratic Splines (RQS) as the transform $f$.

\subsection*{Training and Inference Procedure}
% During training, we pass a batch of $n$ conditions $\mathbf{V} = (\mathbf{v}_1, \mathbf{v}_2, ..., \mathbf{v}_n)$ to the four neural networks and obtain a batch of $n$ base distributions and $n$ RQS flows. We then sample from the $n$ base distributions, one from each base distribution, and use the RQS flows to calculate the $n$ losses corresponding to the batch of $n$ conditions. The overall loss $\mathcal{L}$ is taken as the sum of these $n$ individual losses. The individual loss function used in the training procedure is the KL divergence between the proposal $q(.|v_r;\bm{\theta})$ and the target $p(.|v_r)$ i.e., the true conditional distribution where $r = 1,2,...,n$. The overall loss is given by

The loss function used in the training procedure is the expected value of the KL divergence between the proposal $q(\phi_{i,j}|\mathbf{v}_{i,j};\bm{\theta})$ and the target $p(\phi_{i,j}|\mathbf{v}_{i,j})$ i.e., the true conditional distribution over all possible values of the condition vector $\mathbf{v}_{i,j}$ during training. The first four components of $\mathbf{v}_{i,j}$ lie in the interval  $[0,2\pi]$ and the last component $T \in [0.13, 2.05]$.   We will, therefore, sample $\mathbf{v}_{i,j}$ from $p_v(\mathbf{v}_{i,j}) = $ Unif$([0,2\pi]^4 \times [0.13, 2.05])$ to calculate the expectation.
\begin{equation}
\begin{split}
    \mathcal{L} &= \mathbb{E}_{\mathbf{v}_{i,j} \sim p_v(\mathbf{v}_{i,j})} \left[ \mathbb{E}_{z \sim p_Z(z|\mathbf{v}_{i,j};\bm{\theta}_B)}\left[\log{q(\phi_{i,j} | \mathbf{v}_{i,j} ; \bm{\theta}) - \log{p(\phi_{i,j} | \mathbf{v}_{i,j})}} \right] \right] \\
    % &= \mathbb{E}_{\mathbf{v}_{i,j} \sim p_v(\mathbf{v}_{i,j})} [ \mathbb{E}_{z \sim p_Z(z|\mathbf{v}_{i,j};\bm{\theta}_B)} [\log{p_Z(z|\mathbf{v}_{i,j};\bm{\theta}_B)} + \log{|\det{J_f(z| \mathbf{v}_{i,j};\bm{\theta}_R)}|^{-1}} \\
    % &- \log{p \left(f(z ; \bm{\theta}_R) | \mathbf{v}_{i,j} \right)} ] ]
\end{split}
\end{equation}
The Monte Carlo approximation can be used to estimate the above expectation as follows
\begin{multline}
    \mathcal{L} \approx \frac{1}{n} \cdot \frac{1}{N} \sum_{r=1}^n \sum_{k=1}^N [\log{p_Z(z_k|(\mathbf{v}_{i,j})_r;\bm{\theta}_B)} + \log{|\det{J_f(z_k| (\mathbf{v}_{i,j})_r;\bm{\theta}_R)}|^{-1}} \\
    - \log{p \left(f(z_k ; \bm{\theta}_R) | (\mathbf{v}_{i,j})_r \right)} ]
\end{multline}
We sample from $p_v(\mathbf{v}_{i,j})$ to generate 10,000 samples of $\mathbf{v}_{i,j}$ and use this as the training set. We use the Adam optimizer with default hyperparameters and a cosine decay schedule for the learning rate with an initial learning rate of 0.0005 and 20,000 decay steps. We perform validation by calculating the average acceptance rate for 24 random temperatures in the range [0.13, 2.05] and stop training when we achieve an average acceptance rate of around 85\%. Our model gets trained quickly, taking less than an hour to train on a low-end single GPU machine.

The procedure for MCMC sampling using PBMG-XY is briefed in the algorithm given below. Here, $\mathbf{V}_g = [\mathbf{v}_{i,j}]_{(i,j) \in g}$.

\SetKwInput{KwInput}{Initialize}                % Set the Input
\SetKwInput{KwOutput}{Return}              % set the Output
% \begin{algorithm}[H]
%     \KwInput{$\bm{\phi}$ as $\bm{\phi}^{(0)}$}
%     \For {$t = 0$ to $Num-1$}{
%         Calculate $\mathbf{V}_{g_0}$ at timestep $t$ \\
%         Condition $K_{g_0}$ on $\mathbf{V}_{g_0}$ and obtain from it $\phi^{(t+1)}_{i,j} \forall (i,j) \in g_0$ \\
%         Calculate $\mathbf{V}_{g_1}$at timestep $t$ \\
%         Condition $K_{g_1}$ on $\mathbf{V}_{g_1}$ and obtain from it $\phi^{(t+1)}_{i,j} \forall (i,j) \in g_1$
%     }
%     \KwOutput{${\bm{\phi}^{(0)}, \bm{\phi}^{(1)}, ..., \bm{\phi}^{(Num)}}$}
%     \caption{MCMC sampling procedure using PBMG-XY} \label{alg1}
% \end{algorithm}

\section*{Appendix B: PBMG for $\phi^4$ model }
\label{phi4}

\subsection*{Modeling the Proposal Distribution: PBMG-\texorpdfstring{$\phi^4$}{ } }

We construct the proposal distribution for the scalar $\phi^4$ theory by using a Gaussian Mixture Model with six Gaussian components. 
The proposal distribution parameterized by $\bm{\theta}$, can be written as 
\begin{align}
 q(\phi_{i,j}|\mathbf{v}_{i,j}; \bm{\theta}) = \sum_{k=1}^6 \pi_k(\mathbf{v}_{i,j}; \bm{\theta}_k) \mathcal{N}(\phi_{i,j}| \mu_k(\mathbf{v}_{i,j}; \bm{\theta}_k), \sigma_k(\mathbf{v}_{i,j}; \bm{\theta}_k))
\end{align}
where $\mu_k$, $\sigma_k$ $\pi_k$ are the mean, standard deviation and mixing coefficients of the $k^{th}$ Gaussian distribution, and $\bm{\theta} = \{\bm{\theta}_k \}_{k = 1}^{6}$. The parameters $\mu_k$, $\sigma_k$ $\pi_k$ are a function of the condition vector $\mathbf{v}_{i,j}=(\lambda, m^2, \{\phi_{l,m}: (l,m)\in n(i,j)\})$ through a neural network parametrized by $\bm{\theta}_k$ which are learnt using a suitable loss function. The input to the $k^{th}$ neural network is the condition vector $\mathbf{v}_{i,j}$, and the outputs are the parameters $(\mu_k,\log(\sigma_k), \pi_k)$. %We apply the activation function ReLU(alpha = 1, max value = 1, threshold = 0) to $(\log(\sigma_k))'$ to obtain $\log(\sigma_k)$

The architectures of all six neural networks are the same, with the only difference lying in the initialization of the network parameters. Each neural network consists of one hidden layer with 500 neurons and a ReLU activation function. The neurons in the final layer use linear activation. The value of $\log(\sigma_k)>1$ is clipped to $1$. Since $\sum_k\pi_k=1$, the networks output logit values that are converted to $\pi_k$ by applying softmax. 

\subsection*{Training and Inference Procedure}
The training procedure of PBMG-$\phi^4$ is similar to that of PBMG-XY. The loss function used in the training procedure is the expected value of the KL divergence between the proposal $q(\phi_{i,j}|\mathbf{v}_{i,j};\bm{\theta})$ and the target $p(\phi_{i,j}|\mathbf{v}_{i,j})$ i.e., the true conditional distribution over all possible values of the condition vector $\mathbf{v}_{i,j}$ during training, along with an $L_2$ regularization term. The effect of the regularization term is that the mixing coefficients remain close to each other, which, in turn, facilitates accurate training. We train our model for the following range of the parameters: $\lambda \in [2.5, 15]$, $m^2 \in [-8, 0]$ and $\kappa_{i,j} \in [0,3]$. We will, therefore, sample $\mathbf{v}_{i,j}$ from $p_v(\mathbf{v}_{i,j}) = $ Unif$([2.5,15] \times [-8, 0] \times [0,3])$ to calculate the expectation. Here, $\bm{\pi}(\mathbf{v}_{i,j}; \bm{\theta}) = [\pi_k(\mathbf{v}_{i,j}; \bm{\theta}_k)]_{k=1}^6$ and $\|.\|$ represents the $L_2$ norm.

\begin{equation}
    \mathcal{L} = \mathbb{E}_{\mathbf{v}_{i,j} \sim p_v(\mathbf{v}_{i,j})}\left[\mathbb{E}_{\phi_{i,j} \sim q(\phi_{i,j}|\mathbf{v}_{i,j}; \bm{\theta})}\left[\log{\frac{q(\phi_{i,j}|\mathbf{v}_{i,j}; \bm{\theta})}{p\big(\phi_{i,j}|\mathbf{v}_{i,j})}} \right] + \|\bm{\pi}(\mathbf{v}_{i,j}; \bm{\theta}) \| \right]
\end{equation}
And the Monte Carlo approximation to the above expression is
\begin{equation}
    \mathcal{L} \approx \frac{1}{n} \sum_{r=1}^n \left[ \frac{1}{N}\sum_{k=1}^N\left[\log q((\phi_{i,j})_k|(\mathbf{v}_{i,j})_r; \bm{\theta}) - \log p\big((\phi_{i,j})_k|(\mathbf{v}_{i,j})_r)\right] + \|\bm{\pi}\left((\mathbf{v}_{i,j})_r; \bm{\theta} \right) \|\right]
    \label{loss_phi}
\end{equation}
We sample from $p_v(\mathbf{v}_{i,j})$ to generate 17,500 samples of $\mathbf{v}_{i,j}$ and use this as the training set. We use the Adam optimizer with a learning rate of 0.0001 and default hyperparameters. We perform validation by calculating the average acceptance rate for 50 random sets of parameters ($\lambda$, $m^2$) where $\lambda \in [2.5, 15]$ and $m^2 \in [-8,0]$, and stop training when we achieve an average acceptance rate of around 98\%. PBMG-$\phi^4$ also gets trained very quickly taking only a few minutes to train on a low-end single GPU machine.

% (make corrections)
% To propose a new field value at the lattice site x, we pass the corresponding conditional vector to the trained GMM. The new value is accepted through the MH accept-reject test. To update the whole lattice,  we first update the black sites together and then update the white sites.
The procedure for MCMC sampling using PBMG-$\phi^4$ is exactly the same as that of PBMG-XY. 

% During inference, however, while calculating $V_{g_k}$ where $k = 0,1$, we take the absolute value of $\kappa_{i,j}$ and sample $\phi_{i,j}$ from the resulting GMM. We then consider $-\phi_{i,j}$ as the final sampled value and proceed with inference. This procedure is correct because,
% \begin{equation}
%     S_{loc}(\phi_{i,j},\lambda,m^2,-\kappa_{i,j}) = S_{loc}(-\phi_{i,j},\lambda,m^2,\kappa_{i,j})
% \end{equation}
% \bibliographystyle{JHEP}  % or use another style like unsrt, alpha, etc.
\bibliography{bibb.bib}
\newpage
\end{document}